# A waveguide overloaded cavity kicker for the HLS II longitudinal feedback system*

LI Wu-Bin(栗武斌)[1)]  ZHOU Ze-Ran(周泽然)  SUN Bao-Gen(孙葆根)[2)]  WU Fang-Fang(吴芳芳)

XU Wei(徐卫)  LU Ping(卢平)  YANG Yong-Liang(杨永良)

National Synchrotron Radiation Laboratory, University of Science and Technology of China, Hefei 230029, China

**Abstract:** In the upgrade project of Hefei Light Source (HLS II), a new digital longitudinal bunch-by-bunch feedback system will be developed to suppress the coupled bunch instabilities in the storage ring effectively. We design a new waveguide overloaded cavity longitudinal feedback kicker as the feedback actuator. The beam pipe of the kicker is racetrack shape so as to avoid a transition part to the octagonal vacuum chamber. The central frequency and the bandwidth of the kicker have been simulated and optimized to achieve design goals by the HFSS code. The higher shunt impedance can be obtained by using a nose cone to reduce the feedback power requirement. Before the kicker cavity was installed in the storage ring, a variety of measurements were carried out to check its performance. All these results of simulation and measurement are presented.

**PACS:** 29.20.db, 29.27.Bd    **DOI:**

## 1  Introduction

In the synchrotron light source, an electron storage ring with many bunches is necessary to meet the demand for the high brightness. The electromagnetic field created by these bunches can interact with the surrounding metallic structures generating 'wake fields', which act back on the trailing bunches producing growth of the oscillations. If the growth is stronger than the damping, the longitudinal coupled bunch mode instabilities (CBMIs) occur and the oscillation becomes unstable. In addition, the higher order modes (HOMs) of RF cavities in the storage ring can also cause the longitudinal coupled bunch instabilities.

During the operation of Hefei Light Source (HLS), the longitudinal coupled bunch instabilities were observed, but there were no effective measures to suppress these longitudinal instabilities. It was one of the main limitations of beam intensity. To overcome these instabilities, a new digital longitudinal bunch-by-bunch feedback system will be installed in the storage ring during the upgrade project of Hefei Light Source (HLS II), and the beam intensity will increase to more than 300 mA. The digital longitudinal feedback system consists of a beam position monitor (BPM), a front-end/back-end signal processor, an integrated Gigasample processor (iGp12-45F), two RF power amplifiers and a LFB (longitudinal feedback) kicker. The kicker that is the end of the longitudinal feedback chain transfers the proper energy correction to each bunch. The efficiency of the kicker directly impacts the longitudinal feedback system operation. This paper introduces the development of the longitudinal feedback kicker for the HLS II storage ring.

We choose the waveguide overloaded cavity for our LFB kicker which was developed by DAΦNE at first [1]. Because there are many excellent properties of this type of LFB kicker, such as high shunt impedance and broadband, then it has been adopted at some other accelerator laboratories, for example, PLS, Duke, BESSY II, TLS, etc [2–5].

## 2  Structure of the LFB kicker

The LFB kicker working at its fundamental $TM_{010}$ mode consists of a pillbox cavity, single-ridged overloaded waveguides, and racetrack shape beam pipes which can connect smoothly with the octagonal vacuum chamber of the storage ring without any transition section, and two nose cones to obtain higher shunt impedance. Its internal structure is shown in Fig. 1. The strong coupling of the single-ridged waveguide extends the fundamental mode to enlarge its bandwidth, and can also lead to the remarkable damping of all HOMs of the kicker. It is possible to enlarge its bandwidth more by increasing the number of attached waveguides, namely the number of input/output ports, because of the increasing of the inner wall loss area of the kicker, and two input/output ports are enough for our requirement of the bandwidth. In machine design, the LFB kicker only has three parts, two end-pieces which the waveguides were placed on symmetrically and one cylindrical cavity chamber, as shown in Fig. 2.

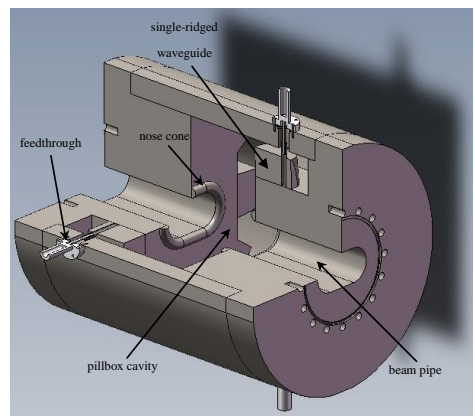

Fig. 1.  The cutaway view of the HLS II LFB kicker.

Received

* Supported by the Natural Science Foundation of China (11175173, 11005105)

1) E-mail: liwubin@mail.ustc.edu.cn

2) E-mail: bgsun@ustc.edu.cn

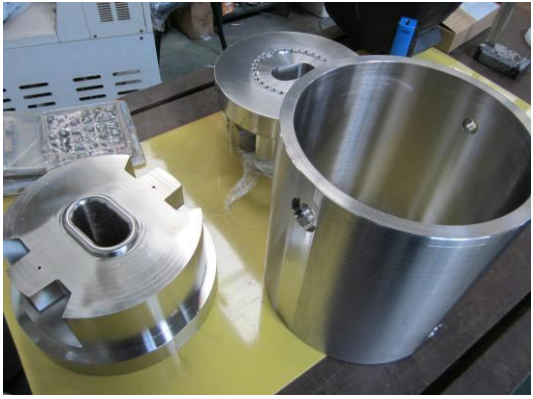

Fig. 2. The three unassembled parts of the LFB kicker.

## 3 Key parameters of the LFB kicker

### 3.1 Central frequency and bandwidth

Let us consider a circulating electron beam in the storage ring with $M$ bunches equally spaced around the ring. Each bunch oscillates at the tune frequency, but there can be some different modes of oscillation, called multi-bunch modes or coupled bunch modes, depending on how each bunch oscillates with respect to the next bunch. Each multi-bunch mode is characterized by a bunch-to-bunch phase shift of

$$\Delta\Phi_n = n\frac{2\pi}{M}, \quad (1)$$

where $n$ is the coupled bunch mode number ($n=0,1,\ldots,M-1$).

In the frequency domain, the CBMIs generate various sidebands around revolution harmonics according to CBMIs theory. These physically positive sideband frequencies are located at [7]

$$f_{p,n,m} = |p \cdot M + n + m \cdot v_s| \cdot f_0, \quad -\infty < p < \infty, \quad (2)$$

where $p$ is an integer, $v_s$ is the synchrotron tune, $f_0$ is the revolution frequency, $m$ is the bunch oscillation mode number associated with the longitudinal oscillation (within bunch mode number, $m=1$ for the dipole mode, $m=2$ for the quadrupole mode, $m=3$ for the sextupole mode, and so on).

If all the buckets are fully filled, namely $M$ is the harmonic number of the RF cavity, the above equation can be written as

$$f_{p,n,m} = |p \cdot f_{RF} + n \cdot f_0 + m \cdot f_s|, \quad -\infty < p < \infty, \quad (3)$$

where $f_{RF}$ is the RF frequency, $f_s$ is the synchrotron frequency.

Since any $f_{RF}/2$ portion of the beam spectrum contains the information of all potential coupled bunch modes and can be used to detect instabilities and measure their amplitude, the minimum bandwidth of the LFB kicker to suppress all CBMIs must be $f_{RF}/2$. Usually the operating frequency range of the LFB kicker is placed on one side of a multiple of $f_{RF}$, thus the central frequency of the LFB kicker can be chosen as $|(p\pm 1/4)| f_R$. For HLS II storage ring, the $f_{RF}$ is 204 MHz, we choose $4.75 f_{RF}$, so the central frequency of the kicker is 969 MHz and the minimum bandwidth is 102 MHz.

The resonant frequency for the pillbox cavity of the LFB kicker, operating in the $TM_{010}$ fundamental mode, is mainly determined by the pillbox cavity radius $R_1$, which is given by

$$f_c = \frac{c}{2\pi} \cdot \frac{2.405}{R_1}. \quad (4)$$

According to this formula, $R_1$ should be 118.5 mm to achieve the resonant frequency of 969 MHz. Additionally, we also must consider the influences of the other internal structure parameters (such as waveguide gap $g$, back cavity height $R_2$, cavity gap $d$, etc) to the physical design of the longitudinal kicker. The final geometry dimensions are shown in Table 1. The simulation and optimization of the parameters of the LFB kicker are carried out by the HFSS code, the central frequency is 968.8 MHz and the bandwidth is 105 MHz.

Table 1. The main geometry dimensions of the LFB kicker

| parameter | value / mm |
|---|---|
| cavity length $L$ | 350 |
| pillbox cavity radius $R_1$ | 114.3 |
| back cavity height $R_2$ | 71.6 |
| pillbox cavity gap $d$ | 94 |
| waveguide gap $g$ | 8.9 |
| nose cone radius $r$ | 6 |

### 3.2 Shunt impedance

The shunt impedance is a main parameter to describe the efficiency of the LFB kicker. For the input power $P$, assuming no power loss in the cable and waveguide, the shunt impedance is given by

$$R_s = \frac{V_{gap}^2}{2P}, \quad (5)$$

where $V_{gap}$ is the cavity gap voltage (kick voltage).

For a fixed input power, a high gap voltage can be obtained with a high shunt impedance to provide enough energy kick to each bunch. The nose cones attaching the edges of the beam pipe, which can concentrate the $E$-fields along the $z$ axis, are adopted to increase the shunt impedance [2, 6].

The electromagnetic field doesn't remain unchanged during the particles traverse the kicker. The transit time factor was introduced to consider the effect of this field's time variation [2]

$$T = \frac{\sin\theta}{\theta}, \quad (6)$$

$$\theta = \frac{\omega d}{2\upsilon} \approx \frac{\pi f d}{c}, \quad (7)$$

where $v$ is the particle velocity, $f$ ($=\omega/2\pi$) is the operating frequency of the power amplifier, $c$ is the velocity of light, and $d$ is the cavity gap size.

Considering the transit time factor, $V_{gap}$ can be obtained through calculating the E-field integration along the kicker central axis by using HFSS field calculator:

$$V_{gap} = \int_0^L E_z(z) e^{-i[\phi_z(z) - 2\pi f z/c]} dz, \quad (8)$$

where $L$ is the length of the kicker.

Then the shunt impedance can be calculated by the Eq. (8), as shown in Fig. 3. The maximum of the shunt impedance is 1860 Ω which appears at 964 MHz. According to Eq. (6) and (7), when frequency is fixed, the transit time factor $T$ increases as the cavity gap size $d$ decreases, and it is possible to get the higher shunt impedance by decreasing the gap size. When $d$ is fixed, the transit time factor decreases as the frequency increases, and this causes the asymmetry of the shunt impedance around the central frequency, the shunt impedance in the high frequency region will be lower than that of the low frequency region. Refer to the approach of PLS to solve this problem [2], we change the ridge of the port structure from flat to round base to obtain the asymmetric $S_{21}$ parameters, shown in Fig. 8. The $|S_{21}|$ values of the $f>f_c$ region are higher than those of the $f<f_c$ region to compensate the transit time factor decay in the high frequency region. As a result of the compensation, the symmetric shunt impedance around the center frequency can be obtained, as shown in Fig. 3.

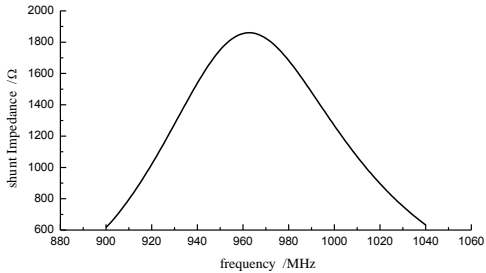

Fig. 3. The shunt impedance of the LFB kicker.

### 3.3 Kick voltage

For longitudinal feedback, the needed maximum kick voltage is given by

$$V_{max} = 2 \frac{1}{\tau_\varepsilon} T_0 (E_0/e) \varepsilon_{max}, \quad (12)$$

where $\tau_\varepsilon$ is the longitudinal damping time constant, $T_0$ is the revolution period, $E_0$ is the beam energy, $e$ is the elementary charge, $\varepsilon_{max}$ is the maximum energy spread.

For the HLS II storage ring, $\tau_\varepsilon = 0.2$ ms, $T_0 = 220$ ns, $E_0 = 800$ MeV, $\varepsilon_{max} = 6 \times 10^{-4}$, the longitudinal maximum kick voltage $V_{max}$ is 422 V. Using the shunt impedance we have obtained, we can estimate the required feedback power in the operating frequency range by Eq. (5) to provide the basis for the choice of the power amplifier.

### 3.4 Higher order modes

The higher order modes (HOMs) of the longitudinal kicker cavity which are under the cut-off frequency of the vacuum chamber of the storage ring can also excite coupled bunch instabilities. The strong waveguide coupling of this kind of longitudinal kicker leads to a remarkable damping of all the cavity HOMs.

Since the octagonal vacuum chamber of the HLS II storage ring is small, it has a higher cut-off frequency (about 4.36 GHz). After further analysis of the beam spectrum, the frequency range which we need to check becomes smaller. The signal of a bunch with Gaussian distribution in time domain is:

$$f(t) = \frac{A}{\sqrt{2\pi}\sigma_\tau} \exp(-\frac{t^2}{2\sigma_\tau^2}), \quad (9)$$

where $\sigma_\tau$ is the bunch length.

The Fourier transform is another Gaussian distribution

$$F(\omega) = \frac{A}{\sqrt{2\pi}} \exp(-\frac{\sigma_\tau^2 \omega^2}{2}). \quad (10)$$

For HLS II, $\sigma_\tau$ is 150 ps. So we have

$$\sigma_f = \frac{1}{2\pi\sigma_\tau} = 1.061\, GHz. \quad (11)$$

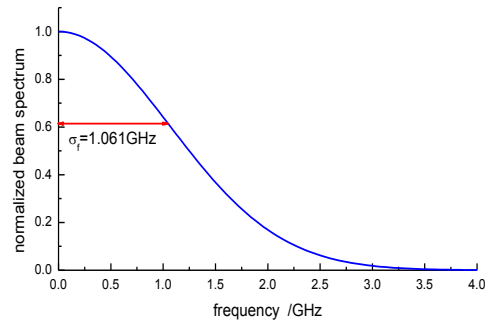

Fig. 4. The normalized beam spectrum.

The normalized beam spectrum is shown in Fig. 4. According to the Fig. 4, it is sufficient to consider the HOMs of the LFB kicker cavity below the 3 GHz which are listed in Table 2, where $Q_L$ is the loaded quality factor. The shunt impedances of these HOMs are less than 4% of that of the fundamental mode, so they are not going to be a significant source of beam instability.

Table 2. The modes of the LFB kicker

| mode | f/GHz | BW/MHz | $Q_L$ | $R_s/\Omega$ | $R_s/Q_L/\Omega$ |
|---|---|---|---|---|---|
| 0 | 0.969 | 105.0 | 9.23 | 1860.0 | 201.52 |
| 1 | 1.562 | 46.7 | 33.5 | 12.1 | 0.36 |
| 2 | 1.705 | 33.8 | 50.4 | 9.3 | 0.18 |
| 3 | 2.242 | 44.1 | 50.8 | 67.8 | 1.33 |
| 4 | 2.418 | 33.5 | 72.2 | 41.2 | 0.57 |

## 4 Measurements of the LFB kicker

The LFB kicker has been manufactured, and a set of measurements must be carried out to check its performance before it was installed in the storage ring. The measurement results are compared with those obtained by simulations.

**4.1 Scattering parameters measurements**

We can assume that the LFB kicker with two input/output ports is a lossless four-port network because the power loss in conductor and the power leakage to the adjacent beam pipes are very small. The reflection coefficient curves ($S_{ii}$) for each port of a well-fabricated LFB kicker should be almost the same. The $S_{ii}$ measured results can be used to test the RF performance of each port. For example, if we want to measure the $S_{11}$, the port 1 should be connected to signal output of the network analyzer and the other 3 ports are terminated with 50Ω, as shown in Fig. 5. The $S_{ii}$ measured results of all four ports are show in Fig. 6. The differences of those curves which mainly are induced by the machining error and the feedthrough connection are so small that they will not cause any problem in the actual operation.

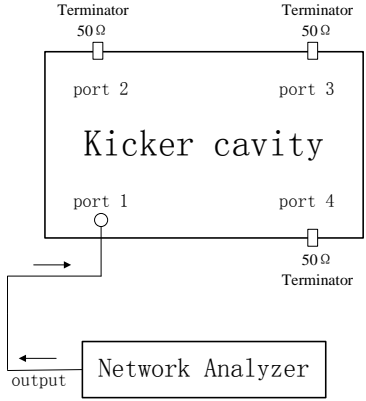

Fig. 5. The setup of reflection measurement of each port.

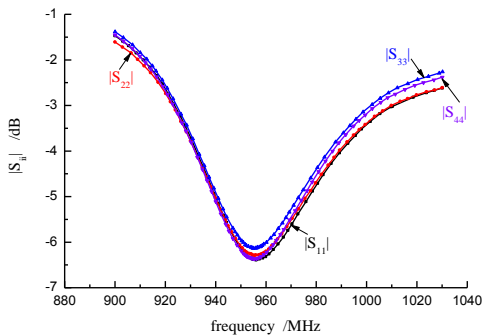

Fig. 6. Measured reflection coefficient of each port.

During the operation of the kicker in the LFB system after being installed on the storage ring, the two input ports on one side are driven in phase with the same amplitude and the two output ports on opposite side are terminated with matched loads. In this operating configuration, the LFB kicker can be treated as a two-port network. The setup for the two-port scattering parameters measurement is show in Fig. 7. The output signal of the network analyzer, after a broadband two-way power splitter, feeds the two input ports of the LFB kicker. Then the signals from the kicker two output ports are recombined to the network analyzer input port.

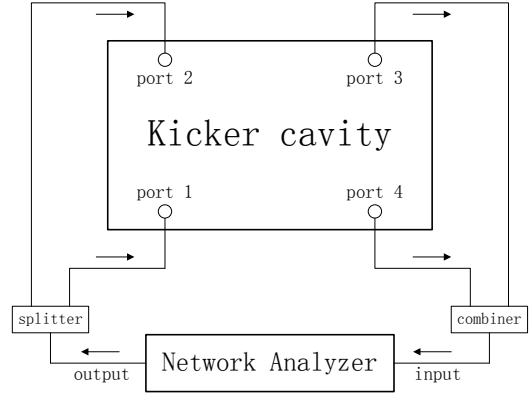

Fig. 7. The setup of reflection and transition measurement when the LFB kicker operates as a two-port network.

The measured results of the reflection and transmission coefficient are compared with the simulation results in Fig. 8. The measured central frequency of the kicker is 954.4 MHz which is about 14.2 MHz (1.49%) lower than the simulated result of 968.6 MHz. The measured 3 dB bandwidth is about 100 MHz which is about 5 MHz (4.76%) lower than the simulated result of 105 MHz. These discrepancies and the low shape distortion of the measured scattering parameter curves were also observed in other LFB kicker cavities [1–3], which are probably due to the mechanical imperfections and the connection of the RF feedthroughs.

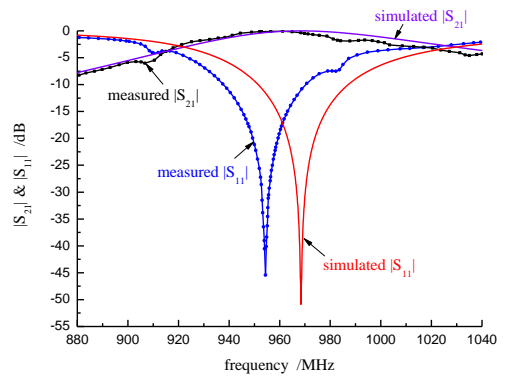

Fig. 8. Comparison of measured and simulated $S$-parameters when the LFB kicker is connected as shown in Fig. 7.

The downshift of the central frequency improves the power transmission performance of the kicker in the low frequency part of the operating frequency range $f_{RF}/2$ (918 MHz~1020 MHz). The values of the $|S_{21}|$ parameter are -3.09 dB and -3.12 dB at 918 MHz and 1020 MHz, respectively, which are very close to the desired -3 dB. The two 100W power amplifiers can supply enough power to produce the excitation voltage which we need.

By means of the network analyzer, the four-port and two-port network scattering parameters have been measured to check the LFB kicker's performance. The simulated and measured results of the main performance parameters of the LFB kicker are listed in the Table 3.

Table 3. The main parameters of the LFB kicker

| parameter | design | simulated | measured |
|---|---|---|---|
| $f_{cent}$ (MHz) | 969.0 | 968.6 | 954.4 |
| BW (MHz) | 102.0 | 105.0 | 100.0 |
| $Q_L$ | 9.5 | 9.22 | 9.64 |
| $R_s$ (Ω) | 1400 | 1860 | N/A |

### 4.2 Electric field on the central axis

It is well known that an inserted dielectric or metal object can change the resonant frequency of a microwave cavity. The distribution of electric and magnetic fields of a resonant cavity can be measured by the perturbation method. This theorem may be stated as [8]:

$$\frac{f - f_0}{f_0} = -\frac{1}{4U} \int_{\Delta V} (\varepsilon_0 |\boldsymbol{E}_0|^2 - \mu_0 |\boldsymbol{H}_0|^2) dv, \quad (13)$$

where $f_0$ is the unperturbed resonant frequency of a cavity, $f$ is its perturbed resonant frequency, $\boldsymbol{E}_0$ and $\boldsymbol{H}_0$ are electric and magnetic fields in the unperturbed cavity, $\Delta V$ is the volume of the small perturbing object, $U$ is the stored RF energy in the cavity, $\varepsilon_0$ and $\mu_0$ are the permittivity and permeability of free space, respectively.

The distribution of electromagnetic fields is able to be obtained by moving the small perturbing object on the selected path and measuring the resonant frequency shift at any position along the path. If the perturbing object is a metal ellipsoid (a oblate spheroid, major axis $l$ and minor axis $a$, $a>l$) and the electric field $\boldsymbol{E}_0$ (magnitude is $E_0$) parallel to the $l$-axis without magnetic field $\boldsymbol{H}_0$, Eq. (13) can be written as [8]

$$\frac{f - f_0}{f_0} = -\frac{\pi d^3}{3U} \frac{(\frac{a^2}{l^2} - 1)^{3/2}}{\sqrt{\frac{a^2}{l^2} - 1} - \arctan\sqrt{\frac{a^2}{l^2} - 1}} \varepsilon_0 E_0^2 \quad (14)$$

$$= -G \varepsilon_0 E_0^2.$$

Under the fundamental mode $TM_{010}$ of the LFB kicker, there are only electric fields $E_0(z)$ along the central $z$-axis. In the measurement experiment of the $E_0(z)$ distribution, a small copper cylindrical rod of length $2l=12$ mm and out diameter $2a=13$ mm is adopted since it is difficult to make an ellipsoid with accurate shape. When $0.68<l/a<18.3$, the resonant frequency shift can be calculated by multiplying a correction factor 3/2 to the R.H.S of the equation of its inscribed ellipsoid [9], so we obtain:

$$E_0(z)^2 = -\frac{2}{3G\varepsilon_0} \frac{f - f_0}{f_0}. \quad (15)$$

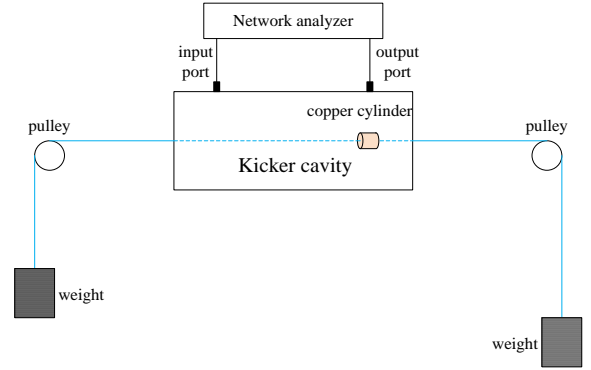

Fig. 9. The experimental setup to measure the electric field distribution along the central $z$-axis.

We move the copper cylinder through the LFB kicker along the central axis using a string and measure the shifted resonant frequency at different position each 5 mm using the setup shown in Fig. 9. The measured and simulated results of the electric field distribution along the central axis are compared in Fig. 10. The measured electric field distribution agrees well with that of the simulation.

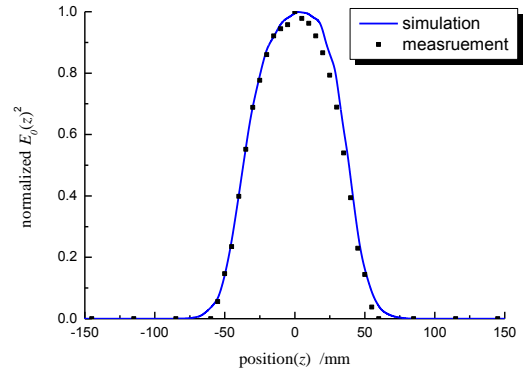

Fig. 10. Comparison of the measured and simulated distribution of the square of the electric field $E_0(z)^2$ along the central $z$-axis.

## 5 Summary

A waveguide overloaded cavity kicker for the HLS II digital longitudinal bunch-by-bunch feedback system has been designed and manufactured. The test experiments of the LFB kicker have been carried out and all these results are in good agreement with the design specifications. In the next future, the LFB kicker will be installed on the storage ring as a key part of the longitudinal feedback system. We expect the new digital longitudinal bunch-by-bunch feedback system can effectively cure the longitudinal coupled bunch instabilities.

# References

1  Gallo A, Boni R, Ghigo et al. Particle Accelerator, 1996, **52**: 95–113
2  Kim Yujong, Kwon M, Huang J et al. IEEE Transactions on Nuclear Science, 2000, **47**(2): 452–467
3  Wu W Z, Kim Y, Li J Y et al. Nuclear Instruments and Methods in Physics Research A, 2011, **632**: 32–42
4  Khan S, Knuth T et al. Proceedings of PAC, New York, 1999, 1147–1149
5  Lau W K, Chang L H, Chen C W et al. Proceedings of PAC, Knoxville, Tennessee, 2005, 949–951
6  XU Wei, HE Duo-Hui et al. Chinese Physics C (HEP & NP), 2013, **37**(3): 037003-1~037003-9
7  Pedersen Flemming. Lecture Notes in Physics, 1994, 425: 269–292
8  Maier L C, Slater J C et al. Journal of Applied Physics, 1952, **23**(1): 68–77
9  Gaspers F, Dome G. Proceedings of the Precision Electromagnetic Measurements Conference, Delft, Netherlands, 1984, 20–24